\documentclass[journal=apchd5,manuscript=letter,layout=twocolumn]{achemso}

\usepackage{chemformula} 
\usepackage[T1]{fontenc} 
\usepackage{hyperref} 
\usepackage{upgreek}
\usepackage{amssymb}

\makeatletter
\let\l@addto@macro\relax
\makeatother
\usepackage[fontsize=11pt]{scrextend}

\author{Lukasz A. Sterczewski}
\affiliation{Jet Propulsion Laboratory, California Institute of Technology, Pasadena, CA 91109, USA}
\alsoaffiliation{Division of Chemistry and Chemical Engineering, California Institute of Technology, Pasadena, CA, 91125, USA}
\email{lukasz.sterczewski@pwr.edu.pl}
\alsoaffiliation{Faculty of Electronics, Photonics and Microsystems, Wroclaw University of Science and Technology, Wroclaw 50370, Poland}
\altaffiliation{These authors contributed equally to this work.}

\author{Tzu-Ling Chen}
\affiliation{Division of Chemistry and Chemical Engineering, California Institute of Technology, Pasadena, CA, 91125, USA}
\altaffiliation{These authors contributed equally to this work.}

\author{Douglas C. Ober}
\affiliation{Division of Chemistry and Chemical Engineering, California Institute of Technology, Pasadena, CA, 91125, USA}
\alsoaffiliation{Jet Propulsion Laboratory, California Institute of Technology, Pasadena, CA 91109, USA}

\author{Charles R. Markus}
\affiliation{Division of Chemistry and Chemical Engineering, California Institute of Technology, Pasadena, CA, 91125, USA}

\author{Chadwick L. Canedy}
\affiliation{Optical Sciences Division, Naval Research Laboratory, Washington, DC 20375, USA}

\author{Igor Vurgaftman}
\affiliation{Optical Sciences Division, Naval Research Laboratory, Washington, DC 20375, USA}

\author{Clifford Frez}
\affiliation{Jet Propulsion Laboratory, California Institute of Technology, Pasadena, CA 91109, USA}

\author{Jerry R. Meyer}
\affiliation{Optical Sciences Division, Naval Research Laboratory, Washington, DC 20375, USA}

\author{Mitchio Okumura}
\affiliation{Division of Chemistry and Chemical Engineering, California Institute of Technology, Pasadena, CA, 91125, USA}
\email{mo@caltech.edu}

\author{Mahmood Bagheri}
\affiliation{Jet Propulsion Laboratory, California Institute of Technology, Pasadena, CA 91109, USA}
\email{mahmood.bagheri@jpl.nasa.gov}

\setkeys{acs}{
abbreviations = true,
}

\title{Cavity-Enhanced Vernier Spectroscopy\\
with a Chip-Scale Mid-Infrared Frequency Comb}

\abbreviations{OFC,ICL,QCL,DCS,VS,ISO,PZT}

\begin{document}




\begin{abstract}
Chip-scale optical frequency combs can provide broadband spectroscopy for diagnosing complex organic molecules. They are also promising as miniaturized laser spectrometers in applications ranging from atmospheric chemistry to geological science and the search for extraterrestrial life. While optical cavities are commonly used to boost sensitivity, it is challenging to realize a compact cavity-enhanced comb-based spectrometer. Here, we apply the Vernier technique to free-running operation of an interband cascade laser frequency comb in a simple linear geometry that performs cavity-enhanced chemical sensing. A centimeter-scale high-finesse cavity simultaneously provides selective mode filtering and enhancement of the path length to  30~meters. As a proof-of-concept, we sense transient open-path releases of ppm-level difluoroethane with 2~ms temporal resolution over a 1~THz optical bandwidth centered at 3.64~$\upmu$m.
\end{abstract} 

\section*{}
Optical frequency combs (OFCs)~\cite{picqueFrequencyCombSpectroscopy2019,fortier201920} have advanced the maturation of broadband high-resolution laser spectroscopy for molecular sensing. The discrete pattern of sharp equidistant lines spanning broad spectral bandwidth that can be metrologically traced to primary frequency standards has advanced fundamental science~\cite{spaun2016continuous} and revolutionized the fields of precision optical metrology~\cite{HanschNobelLecturePassion2006} and chemical sensing~\cite{science.aag1862}. Miniaturization has transformed bulky laboratory-grade sources into compact emitters operating across diverse spectral regions, which allows exploitation of the OFC advantages in portable scenarios~\cite{pasquaziMicrocombsNovelGeneration2018,scalari_-chip_2019, Bao:20}. Electrically-pumped chip-scale OFC sources~\cite{scalari_-chip_2019} based on quantum cascade lasers (QCLs)~\cite{hugi2012mid} and interband cascade lasers (ICLs)~\cite{bagheri_passively_2018, schwarz2019monolithic, sterczewski_interband_2021} are of particular interest due to fundamental molecular absorption bands in the midwave-infrared (mid-IR) spectral region.

A primary factor limiting the compactness and complexity of broadband OFC spectroscopy has been resolution of the individual comb teeth. While this can be accomplished using dispersive Fourier-transform methods~\cite{kawai2020time}, a popular alternative is dual-comb spectroscopy (DCS)~\cite{coddington_dual-comb_2016, sterczewskiThzGas2020,sterczewskiMidinfraredDualcombSpectroscopy2019}. DCS is a mode-resolved Fourier-transform technique that converts optical information into an electrical signal via multi-heterodyne beating~\cite{coddington_dual-comb_2016}. The DCS architecture requires no moving parts and can provide high temporal resolution on the order of nano-~\cite{sterczewskiThzGas2020} to microseconds~\cite{klockeSingleShotSubmicrosecondMidinfrared2018}. Nonetheless, experimental challenges include the requirements for high data throughput, microwave-compatible digitizers, and fast photodetection. Consequently, the hardware resources needed to acquire and process DCS signals (interferograms) can burden the instrument power budget and increase overall complexity.

Furthermore, miniaturization of the instrument can be compromised when high sensitivity (ppm/ppb) is needed~\cite{sterczewskiMidinfraredDualcombSpectroscopy2019}. Such measurements typically require lenghtening of the optical path via a multi-pass cell~\cite{Luo:20} or high-finesse optical cavity~\cite{bernhardt2010cavity, hoghooghi2019broadband, science.aag1862}. Although an optical cavity can increase the effective path length to kilometers, the resonance condition must be also satisfied, which in turn requires locking of the comb and cavity. Finally, DCS strictly requires a mutual coherence time greater than the electrical interferogram repetition rate, which is typically generated by hardware lock loops or shared-cavity dual-comb lasing~\cite{liaoDualcombGenerationSingle2020}. Even then, digital phase post-correction techniques~\cite{burghoffComputationalMultiheterodyneSpectroscopy2016, sterczewskiComputationalCoherentAveraging2019, sterczewski_computational_2019-3} are often needed to accurately recover the mode-resolved spectroscopic information.

\begin{figure*}[!ht]
\includegraphics[width=1\linewidth]{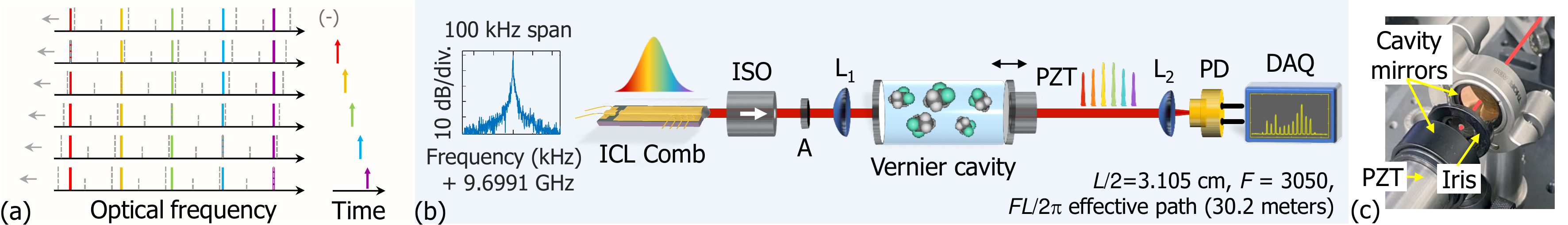}
\caption{(a) Principle of mode-resolved VS, where the dashed lines are cavity modes with spacing and positions that change as the piezo actuator is scanned. Coincidence occurs for only one comb line at a time. Note that the cavity FSR is half the comb repetition rate, therefore every other cavity mode sequentially filters the comb spectrum. (b) Experimental setup of the chip-based VS. (c) Photo of the open-path Vernier cavity.}
\label{fig:introd}
\end{figure*}

An alternative technique that perfectly matches the requirements for compact chip-based OFC spectrometry and circumvents many of the above issues is Vernier spectroscopy (VS)~\cite{gohleFrequencyCombVernier2007a}. At its heart is a tunable high-finesse optical cavity (scanning Fabry-P\'erot interferometer) that simultaneously filters the optical spectrum and enhances the path length. Since VS needs only a single comb source (unlike DCS), the system can operate entirely in free-running mode for significant reduction of the size and complexity. The VS optical cavity functions by providing a secondary scale with respect to the mode spacing of the comb, as in the mechanical Vernier caliper invented in 1631 by Pierre Vernier~\cite{vernierConstructionUsageProprietez1631}. Because the two scales are mismatched, it is possible to get coincidence between one line from each scale, which in this case corresponds to one comb tooth and one cavity mode (Fig.~\ref{fig:introd}a). When applied to tabletop OFC sources~\cite{rutkowskiBroadbandCavityenhancedMolecular2014,khodabakhshFourierTransformVernier2016, khodabakhshMidinfraredContinuousfilteringVernier2017a}, VS has attained impressive temporal resolution and precision. However, the comb's low repetition rate allowed only groups of modes to be filtered instead of single teeth. Also, to allow for very broad optical bandwidth an optical grating mounted on a moving galvo scanner, and synchronized with the piezoelectric transducer (PZT), was needed to disperse higher Vernier orders and prevent groups of comb lines at different wavelengths from reaching the photodetector simultaneously.

In this work, we have used the VS technique to resolve single comb lines in a simple linear geometry. No moving gratings or control loops are needed for spectrally-sparse (yet tunable) semiconductor microcombs, because the Vernier orders can be separated by more than the comb bandwidth while the Vernier cavity filter resolution is much finer than the comb line spacing. Chip-based VS can be seen as complementary to other direct frequency comb techniques like virtual imaging phase array (VIPA)~\cite{nugent-glandorfMidinfraredVirtuallyImaged2012a} spectroscopy, but with the advantage of requiring only a slow single-pixel photodetector rather than a bulky and more expensive array or camera. Beyond the mid-IR, this versatile technique may also be applied to micro-comb sensing systems that cover other challenging spectral regions like the long-wave infrared (LWIR) and even terahertz~\cite{hindleTerahertzGasPhase2019}.

Figure~\ref{fig:introd} illustrates the VS principle~(a) and experimental setup~(b). 
For this prototype, the entire optical layout, including the ICL comb, is contained within a 1.5'$\times$2' area. 
The frequency comb is a GaSb-based ICL active structure~\cite{meyer_interband_2020} processed into a narrow ridge divided into gain and saturable absorber sections~\cite{bagheri_passively_2018} similar to that reported in Ref.~\cite{sterczewskiMidinfraredDualcombSpectroscopy2020}. The ICL is biased in a low-phase-noise regime with <~1~W of electrical power consumption, and emits a 1-THz-spanning spectrum centered at $\nu_c=2750$~cm$^{-1}$ ($\lambda_c=3.636$~$\upmu$m). On careful alignment of a Faraday optical isolator (ISO) and collimation lens, the intermode (repetition rate) beat note centered at $f_\mathrm{rep}=9.6991$~GHz preserves a feedback-free sub-kHz RF linewidth accompanied by $\sim$300~kHz optical line width (estimated from DCS experiments). Narrow optical lines are critical for efficient coupling into the cavity with mode width $\Gamma=1.59$~MHz, since a multi-MHz linewidth broadened by feedback-induced phase noise would effectively preclude mode-resolved measurements and lower the optical transmission by orders of magnitude.

The open-path Vernier cavity shown in Fig.~\ref{fig:introd}c consists of two identical concave mirrors (6 meters radius of curvature) formed by highly-reflective coatings deposited on polished silicon substrates (Los Gatos Research Inc.). The cavity finesse $\mathcal{F}$ was evaluated from the mirror reflectivity data, and independently by a cavity ringdown measurement performed with a single mode laser whose optical frequency is matched to the ICL comb. The ringdown time of $\tau\approx100$~ns yields $\mathcal{F} \approx 3050$ and $R_{1,2}=99.897\%$, which is consistent with the specified mirror reflectivity of $\approx 99.9\%$ at this wavelength.
One of the cavity mirrors is mounted on a piezoelectric transducer (PZT) with maximum displacement $\lambda_\mathrm{max}\approx4$~$\upmu$m that allows scanning the comb spectrum at sub-kHz rate over two Vernier free spectral ranges (FSRs). The scan rate is limited by the inertia of the mirror/piezo assembly and the photon lifetime in the FP resonator~\cite{rutkowskiBroadbandCavityenhancedMolecular2014}.

For efficient coupling into the cavity, the input beam is mode-matched by a lens (L$_1$, $f=500$~mm) that focuses to a point between the mirrors spaced nominally by $L/2\approx3.105$~cm (beam waist $\sim$0.47~mm). Since the corresponding cavity FSR$_c=4.830$~GHz is approximately half the comb repetition rate, every other cavity mode participates in the scanned mode filtering. The cavity \emph{roundtrip} length $L$ was determined by carefully detuning from the perfect-match condition $L_0/2=c/f_\mathrm{rep}=3.093$~cm, such that the measured photodetector signal extended over at least twice the comb's 10-dB bandwidth of 1~THz when the PZT moved by $\lambda_c/2$. This allowed for slight drifts of the comb offset frequency $f_0$ that appeared as a shift of the time-of-arrival (TOA) for the Vernier spectra. The relative comb-cavity mismatch $\epsilon=\Delta L/L_0=2\cdot10^{-3}$ translates to $\Delta L\approx123.7$~$\upmu$m, and yields 
Vernier FSR$_v=c/\Delta L=2.425$~THz, which represents the maximum unaliased optical coverage. For spectroscopy, the roundtrip length $L$ is enhanced~\cite{rutkowskiBroadbandCavityenhancedMolecular2014} by $\mathcal{F}/2\pi\approx486$ to an effective path length $L_\mathrm{eff}=30.2$~m.

A ZnSe objective (L$_2$) focused the cavity-filtered light onto a four-stage thermoelectrically-cooled HgCdTe photodetector (PD) with a transimpedance preamplifier (VIGO PVI-4TE-3.4). Given the current responsivity of the PD $\mathfrak{R}_I(\lambda_c)\approx80$~mA/W at the relevant wavelength, and the transmimpedance gain $K_i\approx200$~kV/A, we derive a voltage responsivity of 16~mV/$\upmu$W. The PD electrical signal was acquired by a 12-bit digitizer (DAQ, Zurich UHFLI) operated in oversampling mode (7~MS/s). From the average peak signal level of tens of mV, we estimate a $\upmu$W-level optical power per comb line. This is <10\% of the light emitted at the ICL facet (8~mW distributed over $\sim$100 lines) because the throughput is severely limited by apertures in the beam path (A) used to improve the slightly elliptical beam profile that would otherwise excite higher order cavity modes. Without beam shaping, the photodetector signal from higher order cavity modes would overlap the fundamental and cause spectral clutter. 

\begin{figure}[!htb]
\includegraphics[width=0.8\linewidth]{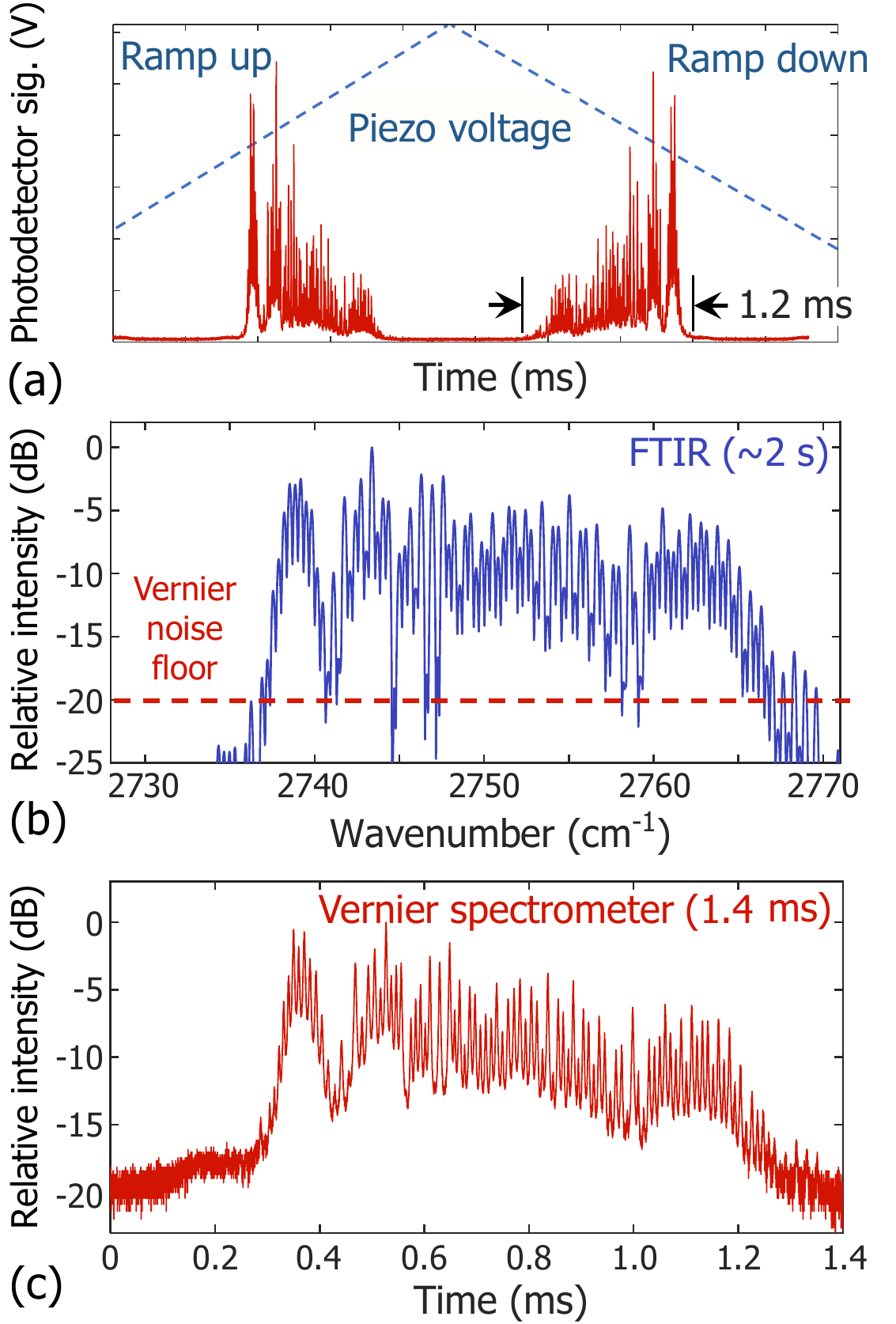}
\caption{(a) Experimental photodetector signal (linear scale) showing both mapping directions. (b) ICL comb spectrum as measured by the FTIR. (c) Vernier spectrum plotted in log scale.}
\label{fig:introd2}
\end{figure}

Figure~\ref{fig:introd2}a shows that when a triangle wave is applied to the PZT, the comb spectrum appears twice in the time-dependent photodetector signal: first in increasing, and next in decreasing frequency order. The sequential mode filtering maps optical frequencies to the time domain with individual comb teeth arriving periodically. Such a signal is referred here to as the Vernier spectrum. The Vernier scan rate of 250~Hz (500~Hz for both directions) was chosen to maximize the temporal resolution while avoiding mechanical resonances and ensuring sufficient temporal separation between comb teeth (due to the cavity finesse and finite ringdown time). To confirm that the entire comb bandwidth is transmitted through the cavity, we also measured the combs's unfiltered optical spectrum with a Fourier transform spectrometer (FTIR, Bruker Vertex 80). We find good agreement between the raw emitted spectrum (Fig.~\ref{fig:introd2}b) and that measured by VS (plotted on the same vertical log scale in Fig.~\ref{fig:introd2}c). However, slight expected discrepancies are observe. First, the FTIR signal requires 1000$\times$ longer to acquire and therefore displays higher signal-to-noise ratio (SNR) than the Vernier scan. Second, the cavity introduces a lineshape convolution effect~\cite{rutkowskiBroadbandCavityenhancedMolecular2014} that appears here as as Lorentzian-profiled lines with full-width at half-maximum (FWHM) $\delta_v=c/\Delta L\mathcal{F}\approx760$~MHz (cavity filter resolution). This convolution lowers the contrast between neighboring lines, although its effect can be removed through deconvolution. In addition, the line intensities in the Vernier signal gradually decrease at higher optical frequencies (up to 2~dB at 2765~cm$^{-1}$) due to progressive line broadening caused by $f_\mathrm{rep}$ fluctuations that lower the comb-cavity coupling efficiency. This effect will be discussed further below.

\begin{figure}[!htb]
\centering
\includegraphics[width=1\linewidth]{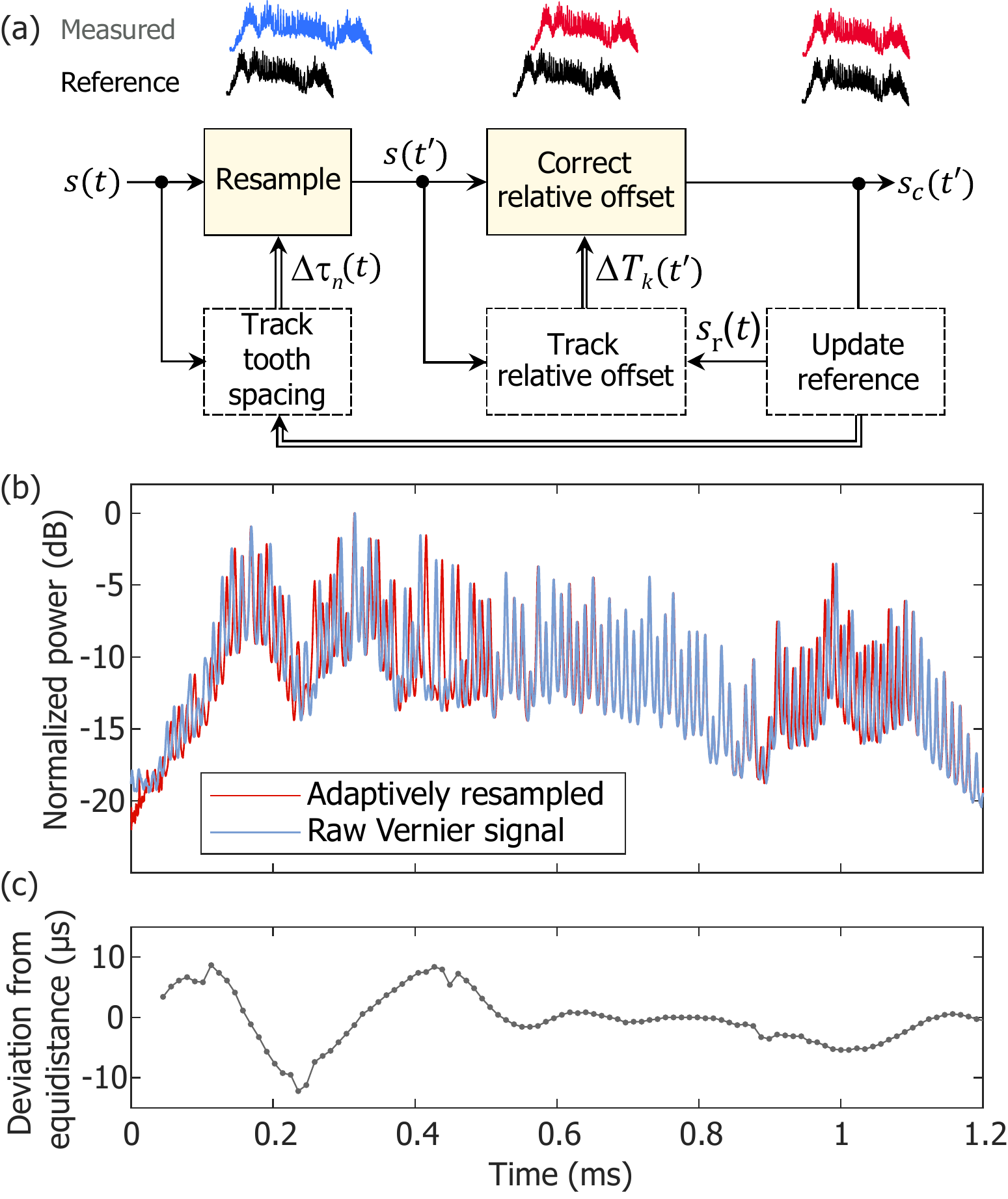}
\caption{(a) Flowchart of the digital correction algorithm. First the time-varying spacing between the peaks is corrected through resampling. Next the relative offset with respect to the reference is eliminated via sample shift. Finally, the reference spectrum (template) is updated to permit operation under rapidly changing conditions. (b) Effect of resampling on the Vernier scan of a free-running ICL comb. (c) Instantaneous deviation from line equidistance.}
\label{fig:resampling}
\end{figure}

In contrast to cavity-locked systems, the VS technique is fully compatible with free-running operation of the laser. If the laser frequency drifts, but remains relatively stable throughout the duration of a sweep ($\sim$MHz drift), Vernier coincidence simply occurs at a different roundtrip length, although with some degradation of the Vernier spectrum. On a short timescale, dynamic optical feedback from the scanned cavity affects the ICL lasing frequency, which induces repetition-rate $f_\mathrm{rep}$ and offset-frequency $f_0$ fluctuations. On a longer scale, laser fluctuations relative to the cavity mode frequency introduce additional frequency noise. 

Time-variations of $f_\mathrm{rep}$ and $f_0$ yield non-equidistant peaks and shifts in the TOA of the Vernier spectrum. This complicates the spectroscopic analysis and precludes coherent averaging. To circumvent this issue, we apply a correction algorithm similar to that developed for unstabilized DCS systems~\cite{sterczewskiComputationalCoherentAveraging2019}, with the difference that the comb signal cannot be described by two global frequencies. Instead, the line spacing relating to $f_\mathrm{rep}$ must be tracked instantaneously (line-to-line)  throughout the scan (Fig.~\ref{fig:resampling}a). Following extraction of the spacing with a robust peak detector, the spectrum is resampled through linear interpolation onto a uniform grid. This resembles adaptive sampling as developed in the context of DCS~\cite{ideguchi2014adaptive}. Figure~\ref{fig:resampling}b and \ref{fig:resampling}c plot results of the line spacing equalization. The locally compressed and expanded spectrum is then corrected through resampling.

\begin{figure*}[!htp]
\centering
\includegraphics[width=0.8\linewidth]{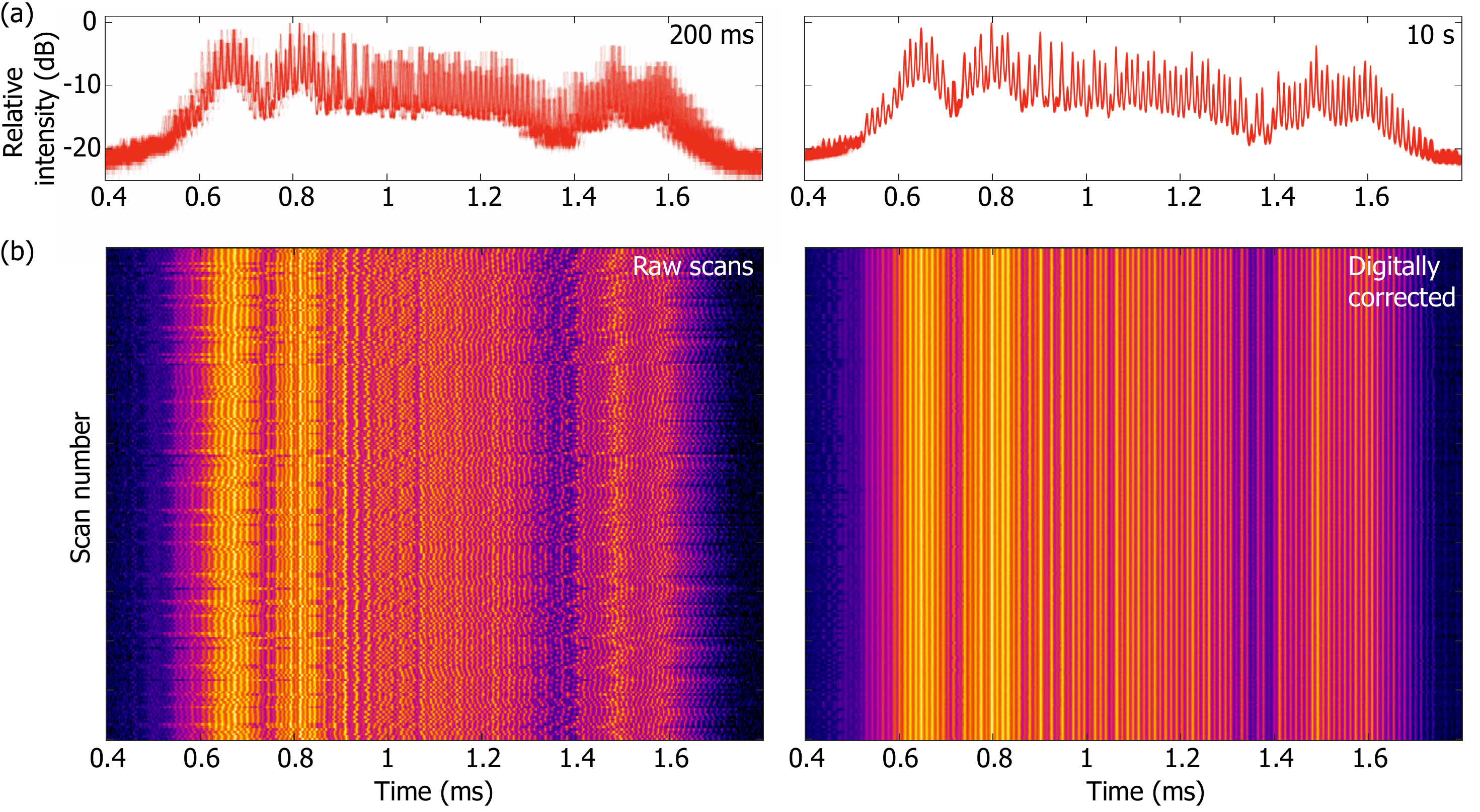}
\caption{(a) Stacked 200 ms of uncorrected Vernier scans (2 ms each, left), compared to a full 10~s acquisition following digital correction (right). (b) Waterfall plot of the data in (a), revealing shifts of the peak positions that are compensated by the algorithm.}
\label{fig:aligmnent}
\end{figure*}

After ensuring equidistant line spacing, we also correct for the relative time offset between the scan frame and reference signal (obtained from an exponential integration of the last $k$ scans/frames). Note that the down-scan must be reversed prior to alignment or simply ignored. The relative time offset $\Delta T_k(t')$ can be obtained by estimating the sample delay from the maximum of the cross-correlation function between the current and reference frame, or by tracking a characteristic peak in a comb spectrum with a unique modal intensity pattern. Figure~\ref{fig:aligmnent} shows the free-running VS data following averaging over 10~seconds with assistance from the digital correction algorithm. Panel~(a) compares an uncorrected 200-ms-long portion (stacked 100 frames) with the full 10-s-long acquisition of 5000 spectra following correction. In the uncorrected data, the initial (low-wavenumber) portion is much sharper and shows less jitter compared to the end-of-scan part. This results from cumulative frequency shift of the $f_\mathrm{rep}$ fluctuations around a fixed point~\cite{newbury_low-noise_2007}. The VS is found to breathe asymmetrically toward higher frequencies, with the frequency (time in the VS scan) jitter increasing uniformly with tooth number.
Despite these imperfections, digital correction fully compensates the drifts. The waterfall representation in Fig.~\ref{fig:aligmnent}b illustrates that digital locking of the fluctuating VS permits power averaging.

\begin{figure*}[!ht]
\centering
\includegraphics[width=0.98\linewidth]{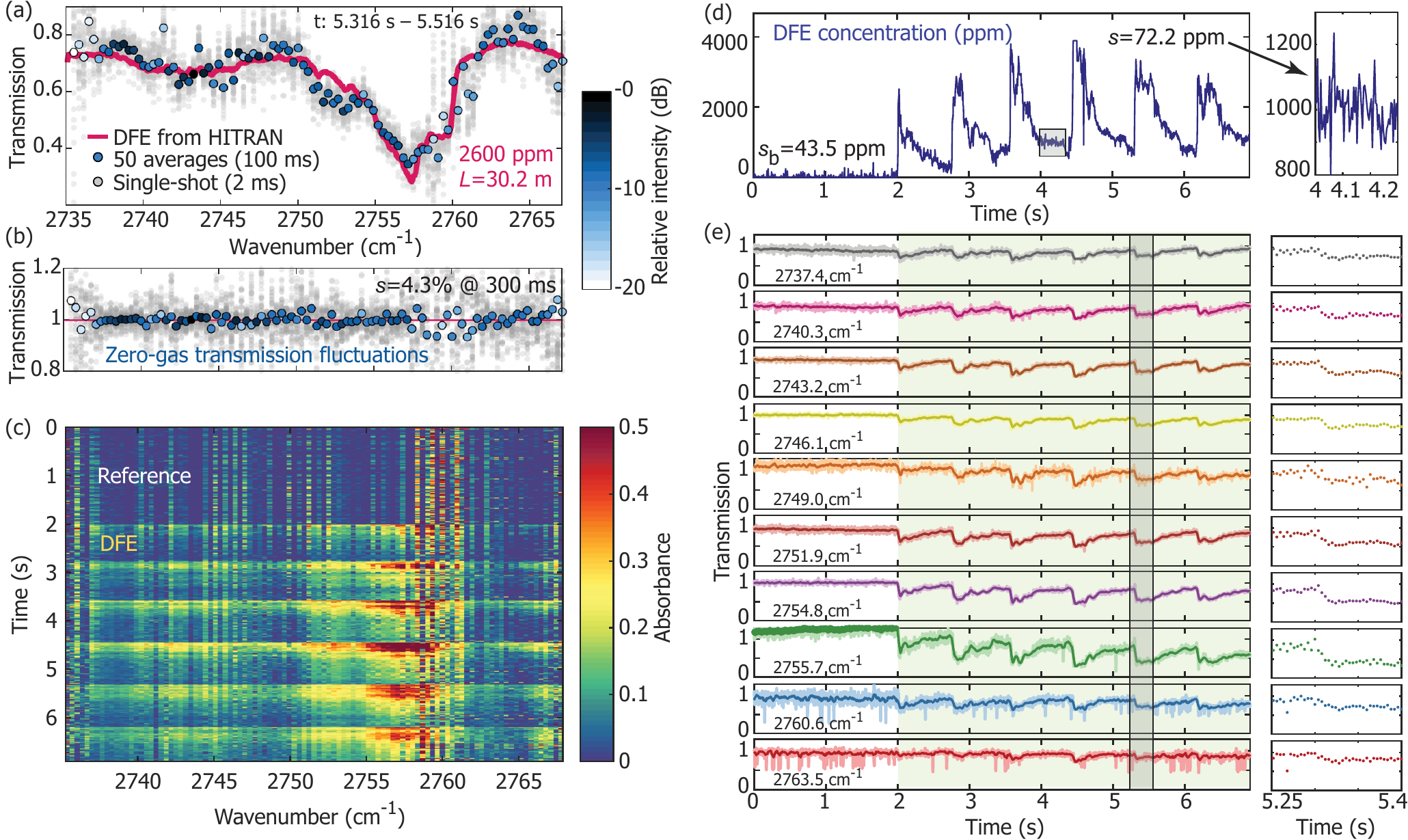}
\caption{(a) VS transmission spectrum (points) along with results from a HITRAN 2016 absorption cross-section model~\cite{kochanovInfraredAbsorptionCrosssections2019} of DFE (red curve). The semi-transparent grey points are single measurements retrieved from the dynamically-evolving signal, while the blue-scale points are averaged over 50 scans for higher precision [darker~=~higher]. (b) Zero-gas (reference) measurement that characterizes the standard deviation of scan-to-scan transmission fluctuations. (c) Spectrogram of the DFE experiment. The vertical features around 2760~cm$^{-1}$, which also appear in the reference spectra, are due to the low-SNR of the relevant comb lines. (d) Time dependenence of the fitted DFE concentration. (e) Transmission of select comb teeth with a zoom on one of the transient absorption events.}
\label{fig:spectroscopy}
\end{figure*}

To demonstrate an application of ICL-comb-based open-path VS to real-time chemical sensing, we used our spectrometer to measure an automotive refrigerant HFC-152a -- 1,1,-Difluoroethane (DFE), which was periodically released in gas phase from a nozzle located $\sim$1~m away from the cavity. The experiment lasting 7~seconds began with measurement of the ambient air (background) followed by quasi-periodic transient gas release events occurring at $\sim$0.8~s intervals starting from $t$=2~s. Figure~\ref{fig:spectroscopy} plots the results of this experiment, during which mode-resolved 1-THz-wide spectra were acquired every 2~ms. After digital scan alignment, we calibrated the optical frequency axis based on the known $f_\mathrm{rep}$ and emission wavelength range. To illustrate the temporal averaging capabilities, Fig.~\ref{fig:spectroscopy}a plots spectra from a later stage of the experiment during quasi-constant DFE concentration. The semi-transparent grey points show modal intensities during a single scan (2~ms) of the dynamically-evolving signal, while the blue-scale points were acquired with longer averaging (100~ms) over 50 scans. The measured spectra show good agreement with HITRAN data ~\cite{kochanovInfraredAbsorptionCrosssections2019} corresponding to a 2600~ppm fit of DFE (red curve).

We see in Fig.~\ref{fig:spectroscopy}a that some line intensities fluctuate more than others, which depends on the differing powers emitted in corresponding teeth of the ICL comb [see Fig.~\ref{fig:introd2}c]. To visualize the non-heteroskedacity (differing precisions) of the data, the fills of the averaged modal intensity points are coloured based on the relative line strength with stronger lines represented by a darker blue. The darker points naturally offer greater precision and deviate less from the model. Nevertheless, a wavy etalon-like structure is superimposed, which is attributable to parasitic reflections at optical interfaces in the system.
To systematically characterize the spectroscopic precision attainable for all comb teeth, Fig.~\ref{fig:spectroscopy}b illustrates the standard deviation of the ratio of consecutive background spectra without the analyte. The standard deviation $s$  of 4.3\% at 300~ms is dominated by outliers with weaker modal intensities. This indicates the need to equalize the spectral envelope of future ICL combs. 

Figure~\ref{fig:spectroscopy}c plots the time-resolved optical spectra acquired in the dynamic gas release experiment. As in Fig.~\ref{fig:spectroscopy}a, it shows intensive broadband absorption caused by the DFE at 2755~cm$^{-1}$ along with a weaker feature at 2742~cm$^{-1}$. In the scanned region, the DFE absorption cross-section reaches $\sim10^{-21}$~cm$^2$/molecule. Although this is two orders of magnitude weaker than at optimal mid-IR wavelengths, it is nonetheless sufficient to demonstrate ppm-level transient detection of the molecule. Figure~\ref{fig:spectroscopy}d plots the fitted DFE concentration as a function of time, which can be visually compared with modal intesities at 10 arbitrary comb frequencies in Fig.~\ref{fig:spectroscopy}e. From the background obtained in the first two seconds, we calculate the standard deviation for concentration $s_\mathrm{b}=43.5$~ppm. Then at $t_0=2$, we see a rapid increase to the 1000s of ppm range, followed by a slow decay as the released gas diffuses in the ambient air (with some oscillations due to turbulent flow of the gas released under pressure). The final 5~s of the scan captures six release events with similar profiles. From the one occurring before $t=4$~s, we retrieve a quasi-steady state concentration of 1000~ppm with standard deviation $s=72.2$~ppm at 2~ms timescale (framed part of Fig.~\ref{fig:spectroscopy}d). The discrepancy between this value ($s$) and that retrieved at the beginning of the experiment ($s_\mathrm{b}$) results from the Rice distribution of the data. At zero gas concentration, the noise is underestimated because the concentration cannot be negative and hence deviates less from the true mean (0). The underestimation factor~\cite{gudbjartsson1995rician} $\sqrt{2-\pi/2}\approx0.655$ renders the value $\hat s
\approx 66$~ppm, which is close to the result  72.2~ppm measured with the analyte. Note that at 2~ms, an ICL comb operating at the near-optimal wavelength ($\lambda\approx3.36$~\textmu m) for DFE may be expected to yield a sub-ppm instantaneous detection limit (1$\sigma$).



\begin{figure}[!t]
\centering
\includegraphics[width=0.92\linewidth]{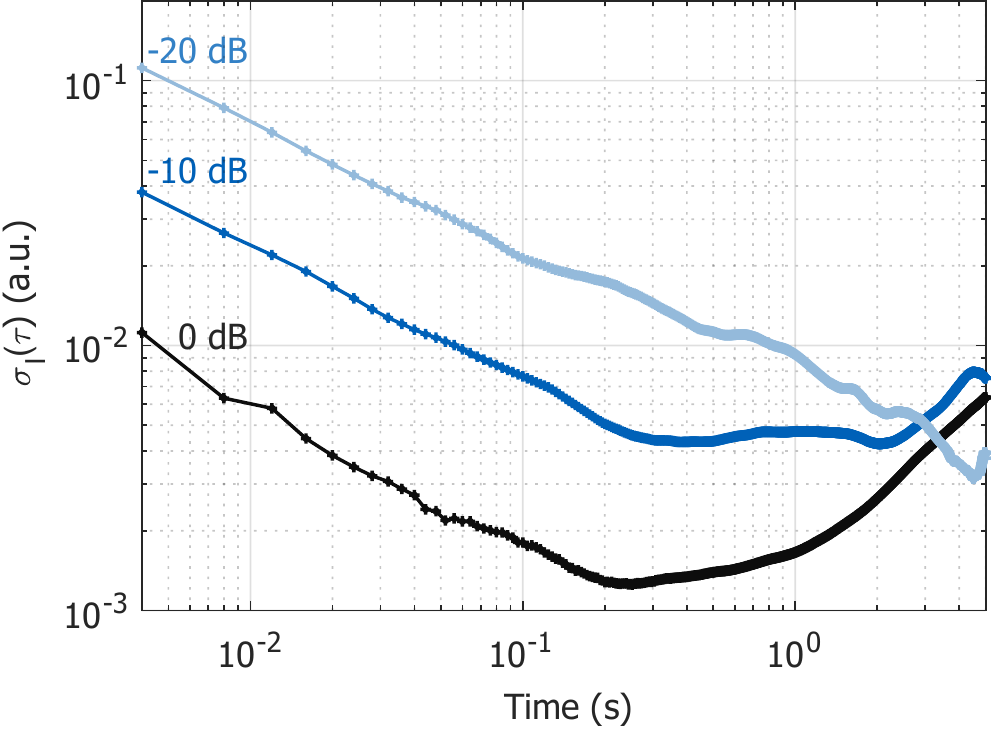}
\caption{Allan deviation plot of the relative intensities for comb lines with three different power levels. The strongest line reaches a minimum in the per-mille precision region, followed by drift after 200~ms.}
\label{fig:Allan}
\end{figure}

To fully characterize the system's capability for averaging comb line intensities, we performed an Allan-Werle deviation analysis~\cite{werle1993limits}. For this we chose the strongest comb line, along with two arbitrary comb lines that were weaker by 10~dB, and 20~dB. These three lines at $\nu_0=2744$~cm$^{-1}$,  $\nu_{-10}=2762$~cm$^{-1}$, and $\nu_{-20}=2735$~cm$^{-1}$ were used to generate Fig.~\ref{fig:Allan}. As expected, for ms timescales a 20~dB decrease in intensity yields 10$\times$ loss in precision. At 2~ms, the precision of the strongest line is already $\sim$1\%, whereas that for the weakest is 10\%. At sub-second timescales, the 0-dB and $-10$-dB curves eventually reach a plateau followed by drift, whereas the $-20$-dB result (dominated by white-noise) continues to improve up to 5~s. A reasonable integration time can be derived from the approximate intercept of all three curves at $\sigma_I\approx0.5$\%, which occurs at $\tau\approx3$~s. The differing trends of the three curves may be attributed to their corresponding levels of white noise. For instance, the $-10$-dB line reaches a plateau when the 0-dB line is already in drift because its averaged noise and drift terms cancel at that point. A more rapid increase of the drift (after 2~s) breaks the plateau balance and causes the $-10$-dB curve to increase as well. The averaging capability is also influenced imperfections in the peak alignment algorithm.

The best-case minimum detectable absorption (MDA) $\sigma_\mathrm{min}$ estimated for the strongest line is 1.3$\times10^{-3}$ at $\tau=240$~ms, which corresponds to a noise-equivalent absorption (NEA) of $\sigma\tau^{1/2}=6.4\times10^{-4}$~Hz$^{-1/2}$, and noise equivalent absorption coefficient (NEAC) of $\mathrm{NEA}/L_\mathrm{eff}=2.1\times 10^{-7}$~cm$^{-1}$Hz$^{-1}$. We emphasize, however, that these numbers do not reflect the broadband sensing capabilities of the demonstrated VS system. We can instead use the background standard deviation of 4.3\% at 300~ms, from which the full-bandwidth NEA is $2.4\times10^{-2}$~Hz$^{-1/2}$, and the NEAC is $7.8\times 10^{-6}$~cm$^{-1}$Hz$^{-1}$. While this is two orders of magnitude lower than for a state-of-the-art VS based on stabilized mid-IR optical parametric oscillators (OPO) on a table top~\cite{khodabakhshMidinfraredContinuousfilteringVernier2017a}, the ICL-comb-based system combines the advantages of compactness, simplicity, mode resolution and free-running operation. Besides tuning the comb spectrum to a near-optimal wavelength $\lambda \approx 3.36$~$\upmu$m, future systems with benefit from improved modal structure that provides a more uniform optical power distribution without spectral gaps~\cite{sterczewski_interband_2021}.

This mode-resolved VS technique opens exciting possibilities for sensitive and robust \emph{in-situ} detection in field campaigns or as a benchtop instrument. Drastically reducing the overall size from this initial demonstration will be straightforward given the limited number of components and the compactness of the ICL comb. Ultimately, all of the optics and the laser could fit within 10 cm without any degradation in performance. This is much smaller than what is possible for FTIR instruments, and is competitive with miniature dual-comb spectrometers.

Future work can further enhance the stability and sensing sensitivity by increasing the finesse or length of the optical cavity. Alternative detection schemes, like cavity ring-down spectroscopy\cite{ORR20111} or one-dimensional spectroscopy\cite{cygan2015one}, can also be implemented using the same experimental setup. The spectral resolution can be improved by tuning the output spectrum via the injection current, so as to fill spectral gaps in the coarse sampling grid defined by the comb's repetition rate~\cite{gianella2020high}. Note from Figs.~\ref{fig:introd2}(b) and \ref{fig:spectroscopy}(a) that under the default operating conditions of the demonstration, the DFE absorption spectrum spanned a region of weak comb emission near 2758~cm$^{-1}$.

While the dual-comb technique is a popular choice for many comb-based spectrometers, mode-resolved Vernier spectroscopy may be viewed as complementary. Its optical path enhancement and selective line filtering eases the requirement for a fast photodetector, microwave digitizers, and two matched comb sources. Despite the lack of truly multiplexed multi-line detection for all the modes simultaneously and constraints on the temporal resolution, advantages include the technique's compatibility with both free-running comb operation and cavity enhancement. These two features allow a relatively simple setup to reach high spectroscopic sensitivity.

In conclusion, a chip-scale mid-IR interband cascade laser frequency comb has been incorporated into a Vernier spectrometer in which the optical cavity enhances the path length, while also selectively filtering the individual comb lines. The ICL comb source with GHz repetition rate and 1~THz of optical bandwidth is well suited to the VS configuration. Under free-running operation in a cavity only 3~cm long, the spectrometer reaches ppm-level sensitivity. The VS also allows dynamic sensing with temporal resolution limited primarily by the cavity scanning rate. With an open-cavity cell, we demonstrate broadband detection of a hydrocarbon in real time with millisecond refresh rate. This prototype system shows the promise of future miniaturized cavity-enhanced spectrometers for chemical sensing. Moreover, the concept may be extended to semiconductor microcombs spanning other spectral regions \emph{e.g.} QCL combs operating in the long-wave infrared or THz.

\subsection*{Notes} The authors declare no conflicts of interest.

\begin{acknowledgement}
This work was supported under and was in part performed at the Jet Propulsion Laboratory (JPL), California Institute
of Technology, under contract with NASA. L. A. Sterczewski’s research was
supported by an appointment to the NASA Postdoctoral Program at JPL, administered by Universities Space Research Association under contract
with NASA. The NRL authors acknowledge support from the Office of Naval Research (ONR). C. R. M. is grateful for support from the Arnold and Mabel Beckman Foundation through the A. O. Beckman Postdoctoral Fellowship. 
\\[5pt]
\noindent\textbf{Funding.} National Aeronautics and Space Agency’s (NASA) PICASSO program (106822 / 811073.02.24.01.85), PDRDF program; Universities Space Research Association (USRA), NASA Postdoctoral Program (NPP).

\end{acknowledgement}

\bibliography{vernier}

\end{document}